\documentclass[]{aastex7}

\begin{document}

\title{Morphology and stellar populations of a candidate ultra-diffuse galaxy in early Euclid and Rubin imaging}

\author[orcid=0000-0003-2473-0369,gname=Aaron,sname=Romanowsky]{Aaron J.\ Romanowsky}
\affiliation{Department of Physics \& Astronomy, San Jos\'e State University, One Washington Square, San Jose, CA 95192, USA}
\affiliation{Department of Astronomy \& Astrophysics, University of California Santa Cruz, 1156 High Street, Santa Cruz, CA 95064, USA}
\email[show]{aaron.romanowsky@sjsu.edu}  

\author[orcid=0000-0003-2876-577X,gname=Yimeng, sname=Tang]{Yimeng Tang} 
\affiliation{Department of Astronomy \& Astrophysics, University of California Santa Cruz, 1156 High Street, Santa Cruz, CA 95064, USA}
\email{ymtang@ucsc.edu}

\author[orcid=0000-0001-9742-3138,gname=Kevin,sname=Bundy]{Kevin A.\ Bundy}
\affiliation{Department of Astronomy \& Astrophysics, University of California Santa Cruz, 1156 High Street, Santa Cruz, CA 95064, USA}
\email{kbundy@ucsc.edu}

\begin{abstract}

We present multi-wavelength imaging and analysis of a low surface brightness (LSB) dwarf galaxy in the Extended Chandra Deep Field South (ECDFS),
SMDG0333094$-$280938, with
particular emphasis on data from the Euclid space telescope and from the Vera C.\ Rubin Observatory.
The galaxy is clumpy and blue, and appears to host globular clusters (GCs), 
suggesting a distance of $\sim$~50--60 Mpc which would make the dwarf an ultra-diffuse galaxy (UDG).
We carry out spectral energy distribution (SED) fitting from the far-ultraviolet to the near-infrared, in order to estimate the galaxy age and metallicity.
We infer a recent peak of star formation that may have led to the formation of the UDG through feedback-driven expansion.
This early analysis illustrates how Euclid and Rubin are poised to identify and characterize many thousands of UDGs and other LSB galaxies in the near future,
including their GCs and stellar populations.

\end{abstract}

\section{Introduction} 

As imaging surveys probe larger areas of the sky at fainter flux levels, populations of LSB galaxies are being increasingly unveiled.
Of particular current interest are
ultra-diffuse galaxies (UDGs), with effective radii
$R_\mathrm{e} \gtrsim$~1.5 kpc and mean SB $\langle \mu_g \rangle \gtrsim 25$~mag~arcsec$^{-2}$
\citep{vanDokkum15}.
The formation histories of UDGs are still uncertain, 
and they display dramatic variations in their dark matter content
-- from very low to very high, with the latter  hosting unusually
abundant GC populations \citep{Forbes24,Buzzo25}.
The largest inventory of candidate UDGs is
SMUDGes (Systematically Measuring Ultra-Diffuse Galaxies;
\citealt{Zaritsky23}), based on the DESI Legacy Imaging surveys \citep{Dey19}.
Here we present 
SMDG0333094$-$280938 at (RA, Decl = 53.2891, $-$28.1606, J2000),
which is the only SMUDGe fully covered
in the overlapping footprints of the first data releases from Euclid and Rubin.
Previous work has been carried out with Euclid on cluster UDGs
\citep{Saifollahi25a,Marleau25a},
but not yet on field UDGs, and there is so far no
publication on any galaxy using Rubin data.

\begin{figure*}[ht!]
\includegraphics[width=\textwidth]{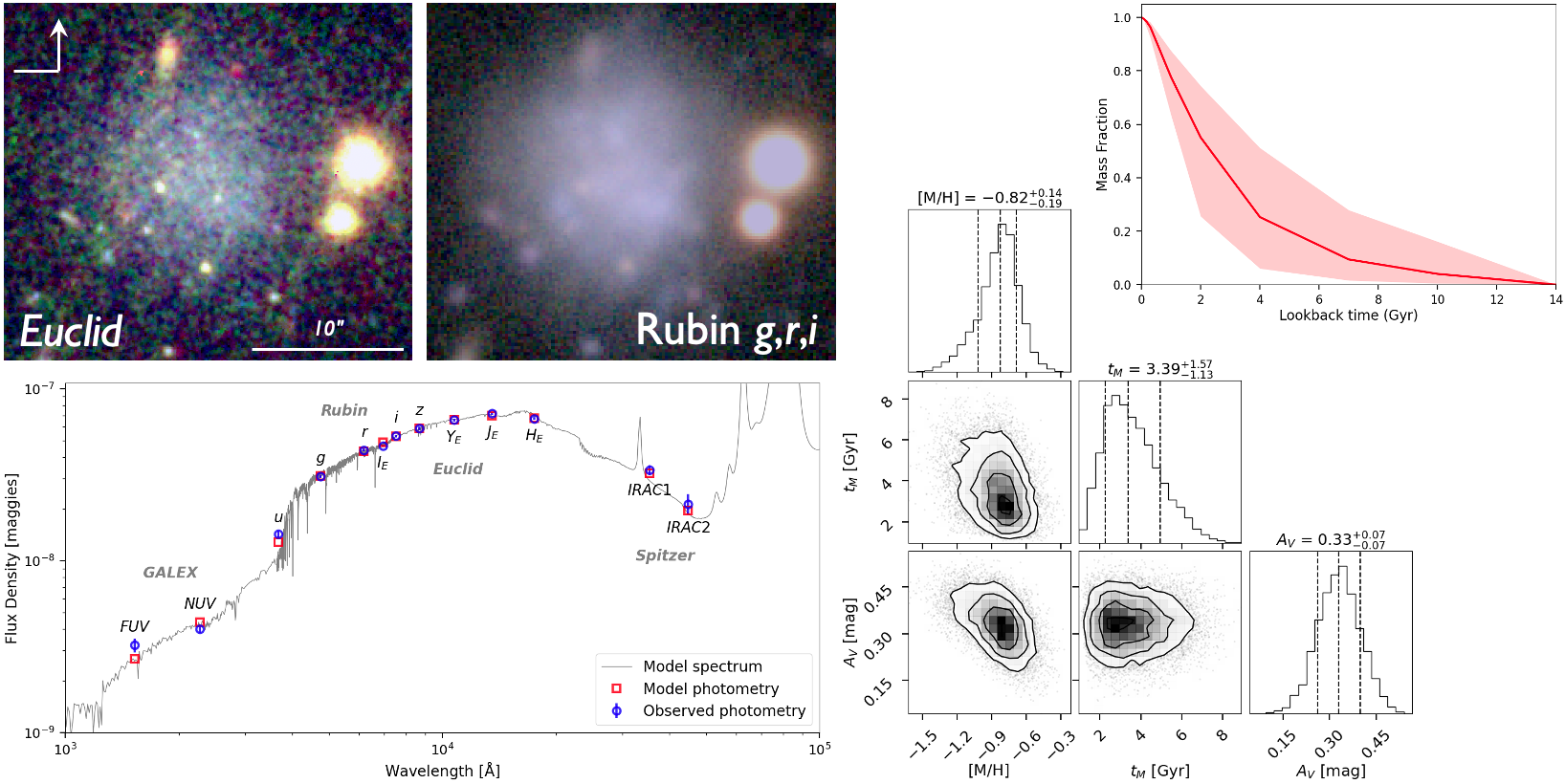}
\caption{The candidate UDG SMDG0333094$-$280938, with  
Euclid image at top left combining visible ($I_{\rm E}$)
and near-infrared light ($Y_{\rm E}, H_{\rm E}$).
The 10$^{\prime\prime}$ scale-bar corresponds to 2.7~kpc if the distance is 55~Mpc.
The Rubin image is at top middle, combining $g, r, i$ filters.
The SED model fit is at lower left, with stellar population results at right,
including posteriors on metallicity, mass-weighted age, dust extinction, and cumulative stellar mass.
\label{fig:general}}
\end{figure*}

\section{Imaging data and morphology} \label{sec:image}

The Euclid imaging
for SMDG0333094$-$280938 is from the Euclid Deep Field Fornax region of the first Quick Data Release (Q1;
\citealt{Q1}).
Figure~\ref{fig:general} shows the Euclid image,
colorized using DS9 (weighting each band equally).
The galaxy has a somewhat irregular shape, including a southwest overdensity suggesting a region of recent star formation.
Some point sources are visible with magnitudes of 
$I_{\rm E} \sim$~25.0--26.5 and colors of $I_{\rm E}-H_{\rm E} \sim$~0--1.0,
consistent with old GCs at a distance of $\sim$~50--60~Mpc
\citep{Hunt25}.
The Legacy-based catalog size of $R_{\rm e} = 6\farcs2$ would correspond to $\sim$~1.5--1.8~kpc at this distance, and 
the SB of $\langle \mu_g \rangle = 24.6$~mag~arcsec$^{-2}$ would qualify this galaxy as a UDG if it
were to quench and fade.

Figure~\ref{fig:general} also shows Rubin LSSTComCam imaging from the Data Products 1 (DP1) release in ECDFS, which has the deepest data in DP1 \citep{DP1},
similar to the full 10-year depth of the Legacy Survey of Space and Time,
with an estimated SB limit of 
$\mu_g \sim 30.3$~mag~arcsec$^{-2}$
and seeing of $\sim 1.1''$.
The image was downloaded from the Rubin Science Platform and colorized using the $g,r,i$ exposures.
The galaxy's color and detailed morphology are readily characterized, in contrast
to the Sloan Digital Sky Survey where UDGs are by definition undetected.
The galaxy appears blue ($g-i \sim 0.6$) and clumpy,
and some of the point sources from Euclid have GC-like optical colors
($g-i \sim 0.85$), although these could be intrinsically bluer: young stellar clusters  reddened by dust
(since it appears unusual for star-forming UDGs to host GCs; \citealt{Jones23}).

We carry out photometry
of this dwarf using the two-dimensional modeling software GALFIT \citep{Peng02}. We adopt a single S\'ersic model,
and a plane sky model,
which provides reasonable fits in all filters (further details in \citealt{Tang25a}).
In addition to the Euclid and Rubin photometry, we also fit archival photometry from GALEX and Spitzer, from 0.15 to 4.5 $\mu$m.
Our resulting sizes and magnitudes are similar to the catalog values from \cite{Zaritsky23}.

\section{Stellar population analysis and discussion}

We carry out SED fitting on the photometric measurements of the galaxy, using
the Bayesian inference code Prospector \citep{Johnson21}. 
To derive a non-parametric star formation history, we adopt 11 time bins and apply the Dirichlet prior. We use linear uniform priors for other parameters.

As shown in Figure~\ref{fig:general},
the SED is fitted adequately and the basic stellar population parameters are well constrained:
metallicity of [M/H]~$= -0.82_{-0.19}^{+0.24}$ dex,
mass-weighted age of $t_M = 3.39_{-1.13}^{+1.57}$~Gyr,
internal dust extinction of
$A_V = 0.33\pm0.07$ mag,
and stellar mass of $\log (M_\star/M_\odot) = 8.02\pm0.09$ for a \cite{Kroupa01} initial mass function and assumed 55~Mpc distance. 
The star formation rate (SFR) rose rapidly over the past $\sim$~4~Gyr to peak $\sim 0.6$~Gyr ago, with a specific SFR of $\sim 3\times 10^{-10}$~yr$^{-1}$, declining to $\sim 5\times10^{-11}$~yr$^{-1}$ over the past $\sim 10$~Myr
(bordering on quiescence).
It may be that there was an unresolved burst of SF extreme enough to cause the galaxy to expand from internal feedback and become a UDG \citep{DiCintio17}.

We note that previous SED modeling of a large sample of SMUDGes found that at young ages, these generally had low metallicities, with a few higher metallicity cases like SMDG0333094$-$280938 \citep{Barbosa20}.
Future work is needed to explore correlations between metallicity variations and GC content for young UDGs.

The metallicity also provides a rough indication of distance, if one assumes that this galaxy follows the standard mass--metallicity relation for dwarfs \citep{Simon19}.
The most likely mass is then $\log(M_\star/M_\odot) \sim 8.7 \pm 1.0$, which although not very constraining does favor a high-mass dwarf.
Scaling to our Prospector results, the favored distance range is then $\sim$~40--400~Mpc, which is
consistent with our initial estimate.

\begin{acknowledgments}
This work has made use of the Euclid Q1 data from the Euclid mission of the European Space Agency (ESA), 2025
hosted at IPAC \citep{https://doi.org/10.26131/irsa601}.
This publication is based in part on proprietary Rubin Observatory Legacy Survey of Space and Time (LSST) data
\citep{https://doi.org/10.71929/rubin/2570308},
and was prepared in accordance with the Rubin Observatory data rights and access policies. 
All authors of this publication meet the requirements for co-authorship of proprietary LSST data.
We thank Rubin staff  for helping navigate this spectacular new dataset.
Other datasets used include
the Bianchi, Conti, Shiao (BCS) Catalog of Unique GALEX Sources \citep{Bianchi2014} hosted at MAST \citep{https://doi.org/10.17909/t9d30c}, and the 
Spitzer Level 2 / post Basic Calibrated Data hosted by IPAC \citep{https://doi.org/10.26131/irsa413}.

This work was supported by NSF grant AST-2308390, by NASA ROSES EGIP24-0037, 
and by the Division of Research and Innovation at San Jos\'e State University (SJSU) 
under Award Number 25-RSG-08-135. 
The content is solely the responsibility of the author(s) and does not necessarily 
represent the official views of SJSU.
\end{acknowledgments}

\facilities{GALEX, Rubin:Simonyi(LSSTComCam, \citet{https://doi.org/10.71929/rubin/2561361}), Euclid, Spitzer}

\bibliography{smudge_rnaas}{}
\bibliographystyle{aasjournalv7}

\end{document}